\documentclass[]{aa}  

\usepackage{graphicx}
\usepackage{txfonts}
\usepackage{hyperref}
\usepackage{natbib} 
\bibpunct{(}{)}{;}{a}{}{,} 

\usepackage{verbatim}
\usepackage{url}
\usepackage{epstopdf}
\usepackage{slashbox}

\begin{document}

   \title{Theoretical investigation on the mass loss impact on asteroseismic grid-based estimates of mass, radius, and age for RGB stars} 

   \subtitle{}

   \author{G. Valle \inst{1,2,3}, M. Dell'Omodarme \inst{3}, P.G. Prada Moroni
     \inst{2,3}, S. Degl'Innocenti \inst{2,3} 
          }
   \titlerunning{Mass loss on RGB stars}
   \authorrunning{Valle, G. et al.}

   \institute{
INAF - Osservatorio Astronomico di Collurania, Via Maggini, I-64100, Teramo, Italy 
\and
 INFN,
 Sezione di Pisa, Largo Pontecorvo 3, I-56127, Pisa, Italy
\and
Dipartimento di Fisica "Enrico Fermi'',
Universit\`a di Pisa, Largo Pontecorvo 3, I-56127, Pisa, Italy
 }

   \offprints{G. Valle, valle@df.unipi.it}

   \date{Received 28/03/2017; accepted //}

  \abstract
{}
   {  
We aim to  perform a theoretical evaluation of the impact of the mass loss indetermination on asteroseismic grid based estimates of masses, radii, and ages of stars in the red giant branch phase (RGB).  
}
{       
We adopted the SCEPtER pipeline
on a grid spanning the mass range [0.8; 1.8] $M_{\sun}$.  As observational constraints, we adopted the star effective temperatures, the metallicity [Fe/H], the  average large frequency spacing $\Delta \nu,$ and the frequency of maximum oscillation power $\nu_{\rm max}$. The mass loss was modelled following a Reimers parametrization with the two different efficiencies $\eta = 0.4$ and $\eta = 0.8$.
}
  {
In the RGB phase, the average random relative error (owing only to observational uncertainty) on mass and age estimates is about 8\% and 30\% respectively.         
The bias in mass and age estimates caused by the adoption of a wrong mass loss parameter in the recovery is minor for the vast majority of the RGB evolution. The biases get larger only after the RGB bump. In the last 2.5\% of the RGB lifetime the error on the mass determination reaches 6.5\% becoming larger than the random error component in this evolutionary phase. The error on the age estimate amounts to 9\%, that is, equal to the random error uncertainty. These results are independent of the stellar metallicity [Fe/H] in the explored range.
 }
{Asteroseismic-based estimates of stellar mass, radius, and age in the RGB phase can be considered mass loss independent within the range ($\eta \in [0.0, 0.8]$) as long as the target is in an evolutionary phase preceding the RGB bump.}


   \maketitle

\section{Introduction}\label{sec:intro}

In recent years, owing to the availability of the high quality asteroseismology data from space born mission such as CoRoT \citep[see e.g.][]{Appourchaux2008,Michel2008,Baglin2009} 
and {\it Kepler} \citep[see e.g.][]{Borucki2010, Gilliland2010}, RGB stars became a target 
of extreme interest \citep[see e.g.][]{Miglio2012MNRAS, Miglio2012b,Stello2015}. 
In fact, the detection of solar-like oscillations in G and K giants  \citep[see e.g.][]{Hekker2009,Mosser2010,Kallinger2010} provided new observational constraints allowing novel approaches in the study of these stars. 
With the  advent of grid techniques relying on classical and asteroseismic constraints for estimation of stellar parameters \citep[see e.g.][]{Stello2009,Quirion2010,Basu2012,Silva2012,scepter1,eta}, it is now possible to obtain fast and precise estimates of the stellar parameters such as mass, radius, and age. 

However, beside the random error due to the uncertainty in the observations, the accuracy of the estimates of these grid methods are prone to systematic biases due to the lack of knowledge of the efficiency of some processes governing the stellar evolution. In some previous works we focussed our attention to the systematic errors arising from the unknown efficiency of the superadiabatic convection, of the microscopic diffusion, and of the convective core overshooting for main sequence stars, either single or in detached binary systems \citep{scepter1, eta, binary}. A demonstration of the perils of blindly relying on grid-based estimates is provided by \citet{TZFor} . That work shows  -- in the framework of fitting a double lined detached binary system -- the theoretical biases resulting on the age estimates from three different sources: intrinsic (i.e. due to the morphology of the stellar tracks that can lead to asymmetric probability of estimates around the true value), caused by the observational errors, and finally due to the choice of the grid of stellar models adopted for the fit.

In this work we focus our theoretical investigation on a potential source of bias in the red giant branch (RGB) phase, that is, the mass loss process whose efficiency is not known with precision. For low mass stars, and thus for old clusters, a relevant amount of mass is  lost during the RGB evolution. This affects the evolutionary fate -- for temperature, path, and duration  -- in the horizontal branch and 
asymptotic giant branch phases, while the RGB evolution is only mildly influenced by this process. In addition, mass loss in the RGB is difficult to measure \citep[see e.g.][]{McDonald2007, Meszaros2009, McDonald2012, Groenewegen2014}.  
Thus RGB mass loss is usually parametrized as a simple function of a free parameter, generally called $\eta$, whose values must be empirically calibrated.  

Recently, a work by  \citet{McDonald2015}  based on 56 Galactic
globular clusters tried to determine the mass loss efficiency in the RGB phase. By adopting two well known mass loss parametrizations \citep{Reimers1975,Schrder2005}\footnote{The difference in the mass loss resulting from the two parametrization is marginal, because the factors added in the more recent one make only a little difference \citep[see][]{Schrder2005}.}, the study poses some constraints on the values of mass loss parameter $\eta$ compatible with observations, suggesting a value of about $\eta \approx 0.5 \pm 0.1$ for the Reimers' formulation. 
Moreover, asteroseismic based constraint on the integrated mass loss was established by \citet{Miglio2012MNRAS} in the old open
clusters NGC6791 and NGC6819, obtaining a mass loss compatible with a value for the parameter $\eta$ in the range [0.1; 0.3] with Reimers' parametrization.

Some  works in the literature have explored the impact of different assumptions on the mass loss efficiency on the estimated stellar characteristics.
\citet{Casagrande2014} performed an estimation of masses and radii of giant stars in the 
Str\"omgren survey for Asteroseismology and Galactic Archaeology (SAGA) in the {\it Kepler} field. They considered scenarios with no mass loss and with maximum mass loss $\eta = 0.4$, finding that the median difference among the inferred parameters is about one half of the random uncertainties owing to the observational errors.   
The work was then extended by \citet{Casagrande2016} analysing, in the SAGA sample, the asteroseismic ages of giant stars (either in the RGB or in the red clump). The study shows a significant impact of mass loss indetermination for red clump stars and very little effect for RGB stars. 
These studies address the impact of mass loss on estimated characteristic of real stars by Bayesian isochrone fitting. As a consequence, the presented results depend on the particular sample analysed. Moreover, they can not avoid the potential contamination of estimates caused by the discrepancies of theoretical models to real stars (e.g. the reliability of scaling relation in this evolutionary phase \citealt{Epstein2014, Gaulme2016}).

The present work adopts a different approach. Rather than dealing with a sample of real stars we preferred a mock catalogue of artificial objects, sampled from the same grid of stellar models used in the recovery procedure. This is the best way to clearly disentangle the effects of the bias source of interest from spurious effects since for all the other stellar parameters models and artificial data are, by construction, in perfect agreement. A pure theoretical approach not only allows a firm estimate of a single source of uncertainty at a time, in this case the mass loss efficiency, but also its dependence on the evolutionary phase along the RGB.  Moreover by means of Monte Carlo simulations it is possible to quantify the very minimum bias which unavoidably affects RGB mass and age estimates owing to the uncertainty in the mass loss efficiency. Such analysis is still lacking in the literature. We aim to fill this gap by presenting a theoretical investigation on the impact of the indetermination in  the mass loss efficiency on asteroseismic grid based  estimates for RGB stars.

\section{Methods}\label{sec:method}

The value of the stellar mass, radius, and age has been determined by means of the SCEPtER pipeline, a maximum-likelihood technique 
relying on a pre-computed grid of stellar models and on a set of observational constraints \citep{scepter1, eta, binary}. 
As observational constraints we adopted the stellar effective temperature, the metallicity [Fe/H], the large frequency spacing $\Delta \nu$, and the frequency of maximum 
oscillation power $\nu_{\rm max}$.
For the reader's convenience, we briefly summarize the technique here.

We let $\cal
S$ be a star for which the observational quantities
 $q^{\cal S} \equiv \{T_{\rm eff, \cal S}, {\rm [Fe/H]}_{\cal S},
\Delta \nu_{\cal S}, \nu_{\rm max, \cal S}\}$ are available. Then we let $\sigma = \{\sigma(T_{\rm
        eff, \cal S}), \sigma({\rm [Fe/H]}_{\cal S}), \sigma(\Delta \nu_{\cal S}),
\sigma(\nu_{\rm max, \cal S})\}$ be the observational uncertainty.

For each point $j$ on the estimation grid of stellar models, 
we define $q^{j} \equiv \{T_{{\rm eff}, j}, {\rm [Fe/H]}_{j}, \Delta \nu_{j},
\nu_{{\rm max}, j}\}$. 
We let $ {\cal L}_j $ be the likelihood function defined as
\begin{equation}
        {\cal L}_j = \left( \prod_{i=1}^4 \frac{1}{\sqrt{2 \pi} \sigma_i} \right)
        \times \exp \left( -\frac{\chi^2}{2} \right)
        \label{eq:lik}
\end{equation}
where
\begin{equation}
        \chi^2 = \sum_{i=1}^4 \left( \frac{q_i^{\cal S} - q_i^j}{\sigma_i} \right)^2.
\end{equation}

The likelihood function is evaluated for each grid point within $3 \sigma$ of
all the variables from $\cal S$. We let ${\cal L}_{\rm max}$ be the maximum value
obtained in this step. The estimated stellar mass, radius,
and age are obtained
by averaging the corresponding quantity of all the models with likelihood
greater than $0.95 \times {\cal L}_{\rm max}$.
Informative priors can be inserted as a multiplicative factor
in Eq.~(\ref{eq:lik}), as a weight attached to the
grid points.

As in \citet{scepter1}, the average large frequency spacing $\Delta \nu$ and
the frequency of maximum 
oscillation power $\nu_{\rm max}$ are obtained using the scaling relations from
the solar values \citep{Ulrich1986, Kjeldsen1995}: 
\begin{eqnarray}\label{eq:dni}
        \frac{\Delta \nu}{\Delta \nu_{\sun}} & = &
        \sqrt{\frac{M/M_{\sun}}{(R/R_{\sun})^3}} \quad ,\\  \frac{\nu_{\rm
                        max}}{\nu_{\rm max, \sun}} & = & \frac{{M/M_{\sun}}}{ (R/R_{\sun})^2
                \sqrt{ T_{\rm eff}/T_{\rm eff, \sun}} }. \label{eq:nimax}
\end{eqnarray}

The validity of these scaling relations in the RGB phase has been questioned in recent years \citep{Epstein2014, Gaulme2016, Viani2017}. Although their reliability poses a severe problem whenever adopted for a comparison with observational data, it is of minor relevance for our aim because we are interested in differential effect with respect to the ideal reference scenario.

\subsection{Stellar model grid}

The grid of models covers the evolution from the zero-age main-sequence (ZAMS) 
up to the tip of the RGB of stars with masses in the range [0.8; 1.8] $M_{\sun}$ (step 0.01 $M_{\sun}$) and initial metallicity [Fe/H] from $-1.3$ dex to 0.55 dex (step 0.05 dex). The grid was computed by means of the FRANEC stellar evolutionary code \citep{scilla2008, tognelli2011}, in the same
configuration adopted to compute the Pisa Stellar
Evolution Data Base\footnote{\url{http://astro.df.unipi.it/stellar-models/}} 
for low-mass stars \citep{database2012, stellar}. 
The initial helium abundance was obtained using the
linear relation $Y = Y_p+\frac{\Delta Y}{\Delta Z} Z$
adopting a primordial $^4$He abundance value $Y_p = 0.2485$ from WMAP
\citep{cyburt04,steigman06,peimbert07a,peimbert07b}, with $\Delta
Y/\Delta Z = 2$ \citep{pagel98,jimenez03,gennaro10}. The models were computed
assuming our solar-scaled mixing-length parameter $\alpha_{\rm
        ml} = 1.74$. Convective core overshooting was
not taken into account. Further details on the input and the related uncertainties are discussed in \citet{cefeidi, incertezze1, incertezze2}.

Besides the reference grid, that is computed neglecting the mass loss, two additional sets of models were computed adopting the widely used mass loss parametrization by \citet{Reimers1975}
\begin{equation}
\dot M = \eta \times 4 \cdot 10^{-13} \times  \frac{L R}{M}  \quad (M_{\sun} {\rm yr}^{-1}),
\label{eq:massloss}  
\end{equation}
where $L$, $R$, and $M$ are the stellar luminosity, radius, and
mass in solar units, and $\eta$ is a fitting parameter, generally assumed lower than one for RGB stars. In the computation we adopted  $\eta$ equal to 0.4 and 0.8, with mass loss process starting at the end of the central hydrogen burning phase.
The values of $\eta$ were chosen in the light of \citet{McDonald2015} and \citet{Miglio2012MNRAS}, as plausible values of the mass loss efficiency.  

The computed stellar grid does not account for the red clump evolution. Although an examination of the impact of the mass loss indetermination on recovered stellar characteristic in this evolutionary phase is interesting, we do not discuss it here for several reasons. First of all we chose to focus our examination on the RGB evolution given the interest of these stars for galactic archaeology studies \citep[e.g.][]{Miglio2013, Stello2015, Anders2017}. Indeed an excellent agreement exists between different stellar evolutionary codes up to the RGB tip, with differences in the evolutionary time scale at the percent \citep[see e.g.][]{Stancliffe2015, testW}. Thus the results presented here can be considered as robust and of general applicability.  
This is no longer the case when post helium flash evolution is considered, because of several important sources of computational indetermination.    
While some stellar codes go smoothly through the helium flash, others do not and zero-age horizontal branch
starting models have to be computed. The adopted code belongs to this second class and relies on the 
procedure described in \citet{database2012}, which is extensively validated but very time consuming. The main problem however is that the stellar evolution in the central helium burning
stage is still more uncertain (e.g. the treatment of core
overshooting and semiconvection; the suppression or not of
the breathing pulses). Different algorithmic choices can easily lead to a difference of dozens of percentage points in the helium burning time scale. Obviously, this uncertainty propagates into the recovery of the mass loss impact, thus confining the applicability of the results only to the specific adopted code and configuration.

\section{Intrinsic accuracy and precision of the stellar parameter estimation}
\label{sec:res-random}

As a preliminary step, we quantified the random error component on the recovered mass, radius, and age  on a synthetic dataset obtained
by sampling $N = 100\,000$ artificial stars from the same standard estimation
grid of stellar models used in the recovery procedure itself and adding
a Gaussian noise in all the observed quantities to
each of them.
Thus this first step adopted the same procedure as that described in \citet{scepter1,eta} when the reference grid is adopted both for the artificial stars generation and for the recovery.
We adopted as standard 
deviation of observational quantities 1.5\% in $\Delta \nu$, 2.2\% in 
$\nu_{\rm max}$, 100 K in $T_{\rm eff}$, and 0.1 dex in [Fe/H]. 
The errors on the asteroseismic quantities were chosen 
considering the values quoted in SAGA of 0.7\% and 1.7\%  on $\Delta \nu$ and 
$\nu_{\rm max}$ \citep{Casagrande2014} and those in the APOKASC catalogue
of 2.2\% and 2.7\% \citep{Pinsonneault2014}.
The sampling stage was performed taking into account the grid evolutionary time steps, so that slow parts of the evolution were sampled more frequently than faster ones.

\begin{figure*}
        \centering
        \includegraphics[height=17cm,angle=-90]{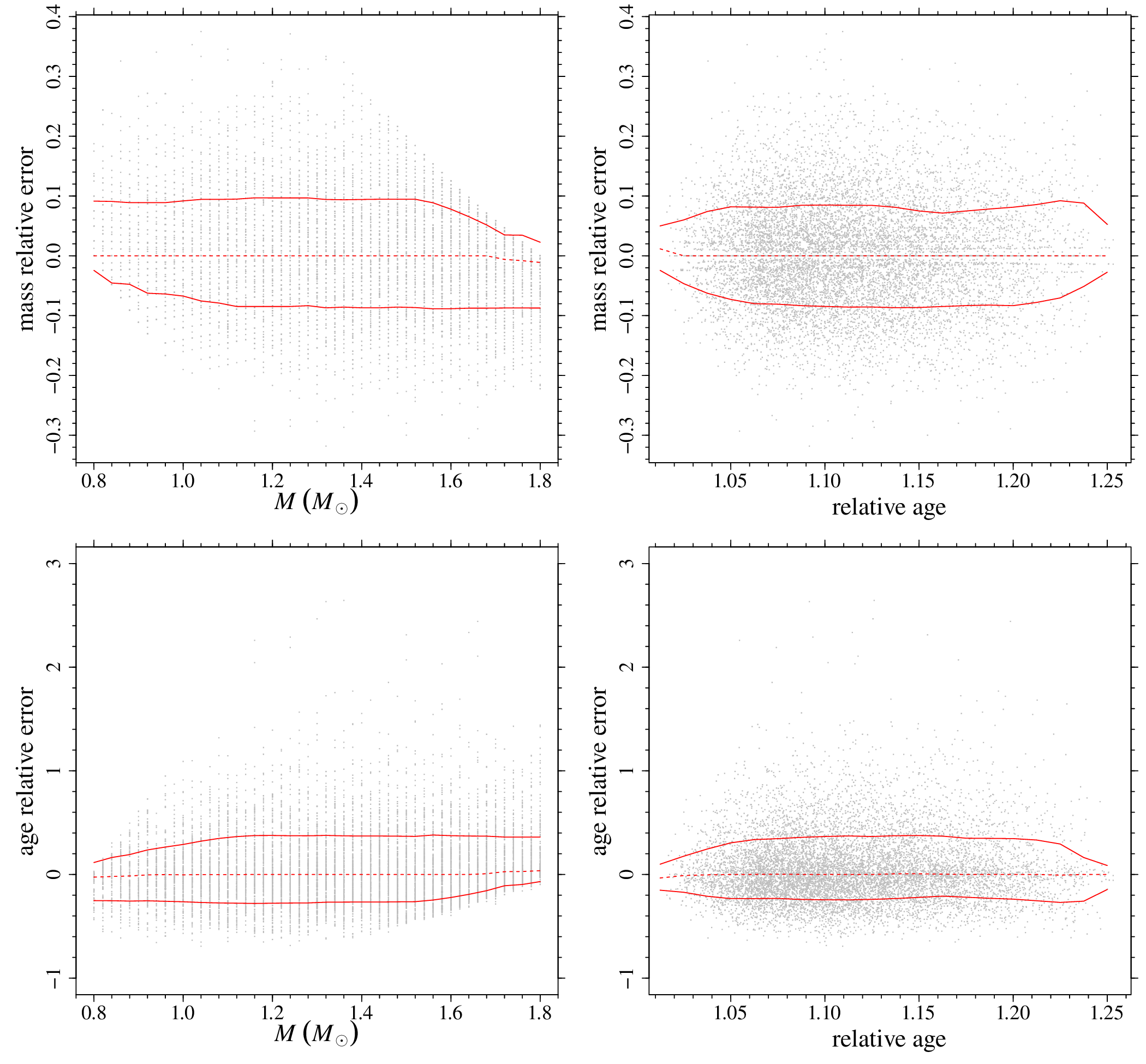}
        \caption{{\it Top row}: {\it left} Monte Carlo relative error -- only due to observational errors -- on the mass as a function of the true mass of the star. The red line shows the relative error $1 \sigma$ envelope. {\it Right} as in the {\it left} panel, but as function of the relative age of the stars. The figure shows only the RGB evolution (see text). {\it Bottom row}: as in the {\it top row}, but for the age relative error.
        }
        \label{fig:envelope-std}
\end{figure*}

The results of the simulations are presented in Fig.~\ref{fig:envelope-std} and Tables \ref{tab:massvsmass} to \ref{tab:agevsage}. The top row of the figure shows the relative error on the reconstructed mass as a function of the true mass of the star (left panel) and of the relative age of the star\footnote{Relative age is defined as the ratio of the current age of the star and the age of the star at the central hydrogen depletion. It is conventionally set to zero at ZAMS.} (right panel). The error envelopes shown in the figure mark the position of the 16th and the 84th quantiles and are constructed as described in \citet{eta}. In this case the figure is focussed on the RGB evolution. The most interesting feature is the increase in the random error as the stars start climbing the RGB. In fact, the mean mass relative error envelope width is about 8\% for relative age higher than 1.05, while it is about 3.5\% at relative age 1.00 (central hydrogen depletion) that is, an increase by more than a factor of two. Regarding the width of the age relative error envelope, its inflation is even more severe, since it increases to a mean level of about 30\% with respect to a value of 12\% at the end of the central hydrogen burning phase.

This behaviour is expected and already reported in the literature \citep{Gai2011, bulge}, and is caused by the narrower packing  of the stellar tracks in this evolutionary phase, which makes the estimate process more difficult than in earlier evolutionary stages. Unlike the results presented in Fig.~\ref{fig:envelope-std}, 
Tables \ref{tab:massvsage} to \ref{tab:agevsage} present the location of the envelope boundaries as a function of the RGB relative age $r$\footnote{RGB relative age is zero at the central hydrogen depletion and 1 at the RGB tip.} and not the total relative age. This choice is motivated by the fact that the relative duration of the  RGB phase  with respect to the MS life is largely variable, mainly owing to the mass of the star. Adopting RGB relative age helps to synchronize the track evolutions and allows better evidencing some slight effects which would otherwise remain concealed.

\section{Impact of assuming a different mass loss in the estimation}
\label{sec:bias}

\begin{figure}
        \centering
        \includegraphics[height=8cm,angle=-90]{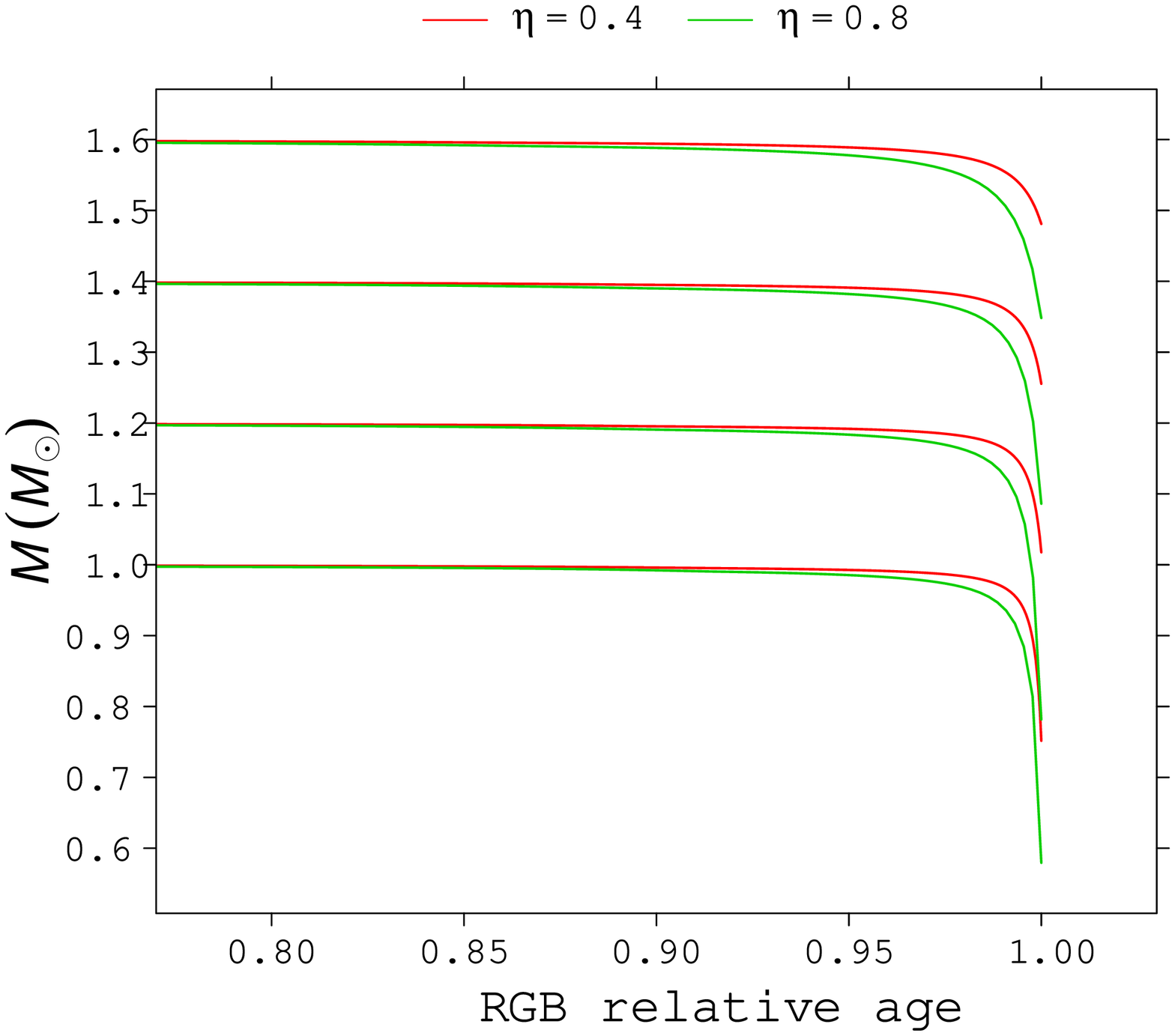}
        \caption{Effect of the mass loss on the mass of the star for $\eta$ = 0.4 (red line) and 0.8 (green line) for stars of masses 1.0, 1.2, 1.4 and 1.6 $M_{\sun}$ at solar metallicity. 
        }
        \label{fig:density-ml}
\end{figure}

Before analysing the impact of adopting in the recovery an $\eta$ value different from the one adopted for the Monte Carlo sampling, we discuss the effect on the total mass value of assuming a mass loss $\eta = 0.4, 0.8$, from the central hydrogen depletion to the RGB tip. 
Figure~\ref{fig:density-ml} shows, for the two assumed $\eta$ and for star of masses 1.0, 1.2, 1.4 and 1.6 $M_{\sun}$ at solar metallicity, the evolution of the stellar mass with the RGB relative age. Following the Reimers' formulation, the effect of mass loss begins to manifest itself after relative age higher than about 0.9 -- when the star is near to the RGB bump -- in agreement with observational results which indicate a very low mass loss during the MS and the first RGB phases. For the 1.0 $M_{\sun}$ model, the final mass is decreased of about 20\% and 40\% for $\eta = 0.4, 0.8$ respectively, while the effects are of about 8\% and 16\% for the 1.6  $M_{\sun}$ model. In fact, the lower the stellar mass, the slower the RGB evolution and the more efficient is mass loss during the RGB phase.
Therefore stars located after the RGB bump are affected by the mass loss, with not precisely determined efficiency.
For ease of comparison with observational data, adopting a reference solar value of $\Delta \nu_{\sun}$ = 135.1 $\mu$Hz, the 90\% of the RGB evolution corresponds to $\Delta \nu$ values in the range [3.9; 6.1] $\mu$Hz for 0.8 $M_{\sun}$ stars (depending on the metallicity) and in the range [1.3; 1.6] $\mu$Hz for 1.8 $M_{\sun}$ stars.

\begin{figure*}
        \centering
        \includegraphics[height=17cm,angle=-90]{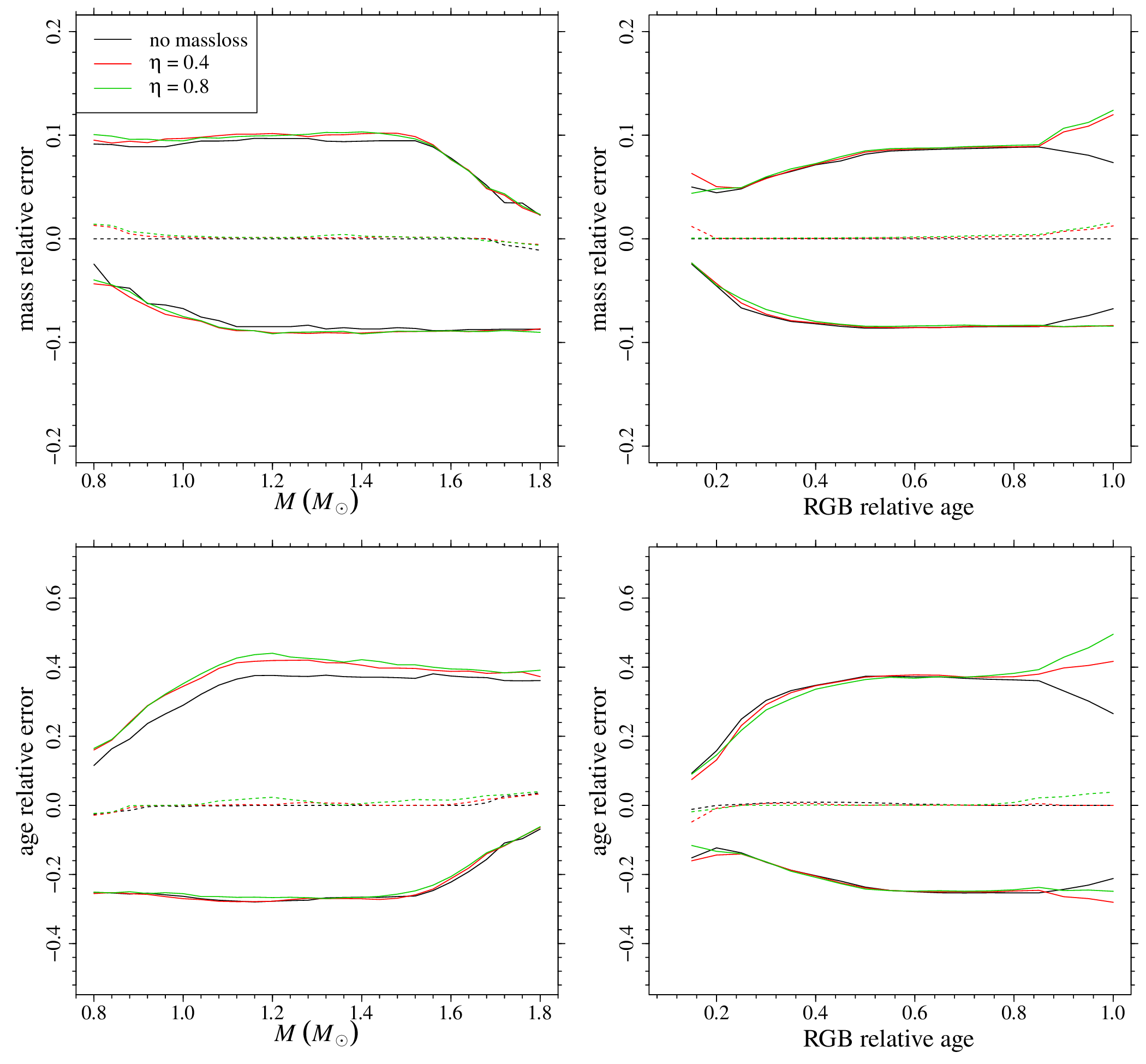}
        \caption{{\it Top row}: {\it left} Monte Carlo $1 \sigma$ envelope on the estimated mass as a function of the true mass of the star. The black solid line shows the relative error $1 \sigma$ envelope for the no mass loss sampling and reconstruction, the red line refers to the sampling from the grid with  $\eta = 0.4$ and reconstruction on the grid with $\eta = 0.0$, while the green one to the sampling from  $\eta = 0.8$ grid and reconstruction on the grid with $\eta = 0.0$. The dotted lines mark the medians.
         {\it Right} as in the {\it left} panel, but as function of the RGB relative age of the stars. {\it Bottom row}: as in the {\it top row}, but for the age relative error.
        }
        \label{fig:envelope-ml}
\end{figure*}

To directly assess
the impact of a wrong assumption of the mass loss efficiency on the estimated stellar parameters, we performed some Monte Carlo simulations. For these tests we sampled $N = 100\,000$ stars from the grids with $\eta = 0.4, 0.8$, perturbed the values as in the previous section to simulate a observational error, and then reconstructed their mass, radius, and age on the standard grid with $\eta = 0.0$. As in the previous section, the sampling takes into account the grid evolutionary time step.
The results are presented in Fig.~\ref{fig:envelope-ml} and in Tables \ref{tab:massvsmass} to \ref{tab:agevsage}.

As expected in the light of the discussion at the beginning of this section, the impact of assuming a different mass loss parameter $\eta$ in the recovery and sampling steps produces negligible differences in the estimates up to about 90\% of the RGB evolution. Therefore the majority of giant stars will be unaffected by this source of uncertainty. This results agree with the analysis presented by \citet{Casagrande2014} and \citet{Casagrande2016} who reported marginal differences in their estimates shifting from a stellar model grid computed with $\eta = 0.0$ to one with $\eta = 0.4$.

From Table~\ref{tab:massvsage} we note only a small bias at $r = 1.0$ in the mass estimate (2\%, that is, about one fifth of the random envelope half width) and a shift towards overestimation  in the position of the upper boundary of the $1 \sigma$ random error envelope which nearly doubles (from about 7\% to 12\%). Moreover, Tables \ref{tab:massvsmass} to \ref{tab:agevsmass} show that the effect of assuming a wrong mass loss vanishes completely for star more massive than about 1.5 $M_{\sun}$, as expected from the inverse dependence of mass loss from the mass in Eq.(\ref{eq:massloss}) and in agreement with the results presented by \citet{Casagrande2016} who claim a negligible effect of the mass loss for stars with mass above 1.7 $M_{\sun}$.

For the stellar age, the results are reassuring because the bias due to a wrong assumption on the mass loss efficiency on the final age estimates is completely negligible for the vast majority of giant stars, supporting the finding by \citet{Casagrande2016}. We detected a small bias in the median age of 4\% for $\eta = 0.8$ at $r = 1.0$, negligible with respect to the random error envelope. Moreover, a shift of the upper error envelope boundary towards overestimation is present (from 27\% to 42\% and to 50\% for $\eta = 0.4$ and $\eta = 0.8$ respectively).

\begin{figure}
        \centering
        \includegraphics[height=8.2cm,angle=-90]{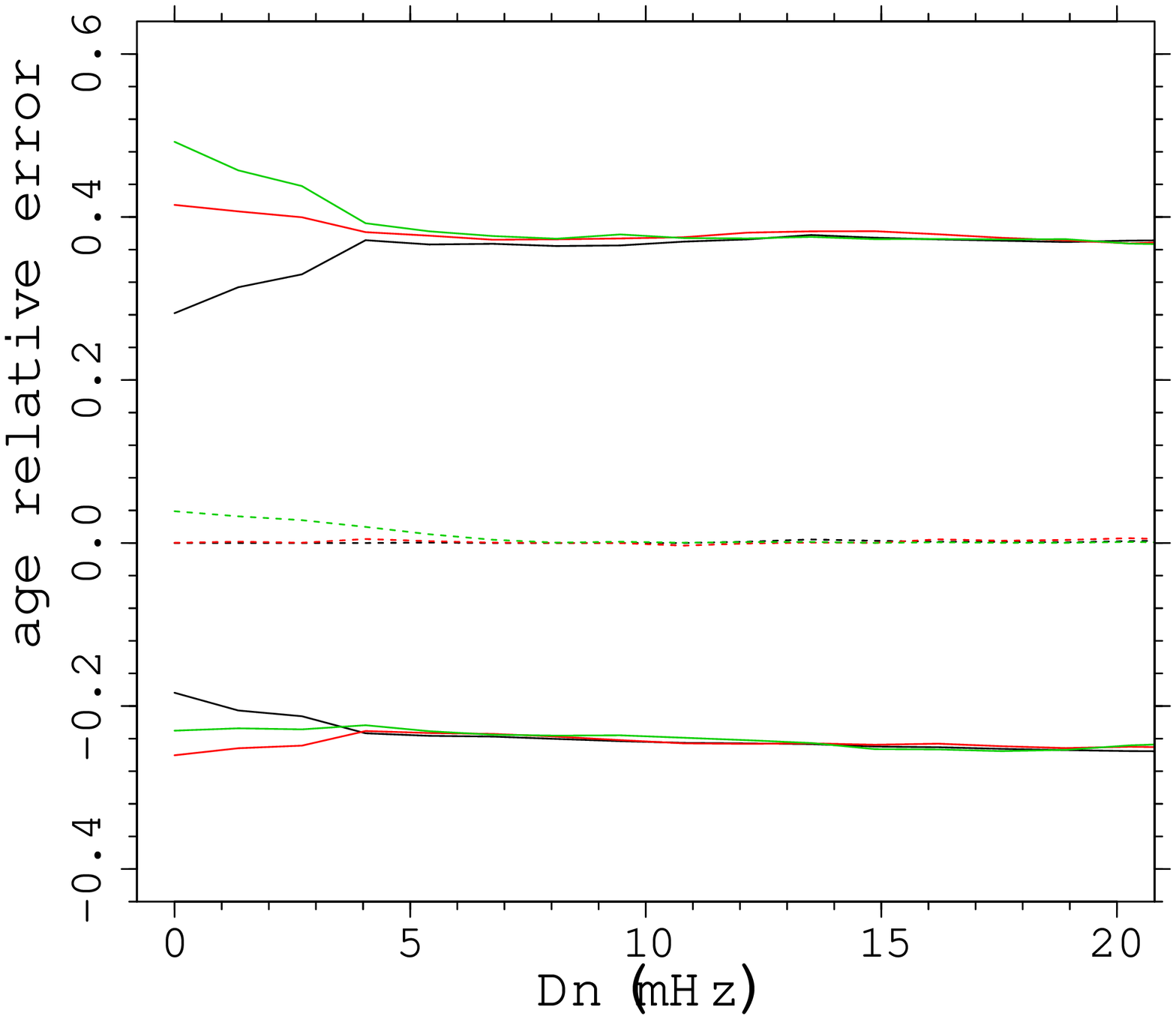}
        \caption{As in Fig.~\ref{fig:envelope-ml}, but as a function of the large frequency separation $\Delta \nu$.
        }
        \label{fig:envelope-Dnu}
\end{figure}

For comparison with observational data, Fig.~\ref{fig:envelope-Dnu} shows the age envelope as a function of the large frequency separation (assuming a solar value of $\Delta \nu_{\sun}$ = 135.1 $\mu$Hz). The impact of the mass loss is negligible for values above 4 $\mu$Hz and is of some relevance below 2 $\mu$Hz. For a rough estimate of the star abundance in these ranges, the values correspond to about 33\% and 10\% of stars in the APOKASC catalogue.

Overall, the results support the conclusion that, as long as stars are located before the RGB bump, the grid-based estimates are safe and robust with respect to the adopted assumptions on mass loss. However the situation could be very different if a star is near to the RGB tip. Restricting the analysis to stars in the terminal 2.5\% of the RGB evolution a direct computation shows a median bias in mass estimate, for the two $\eta$ scenarios, of 6.5\%, which is larger than the random envelope half width in this evolutionary phase. Likewise an underestimation of about 9\%, equal to the random envelope half width, occurs for age estimates.

\begin{figure*}
        \centering
        \includegraphics[height=17cm,angle=-90]{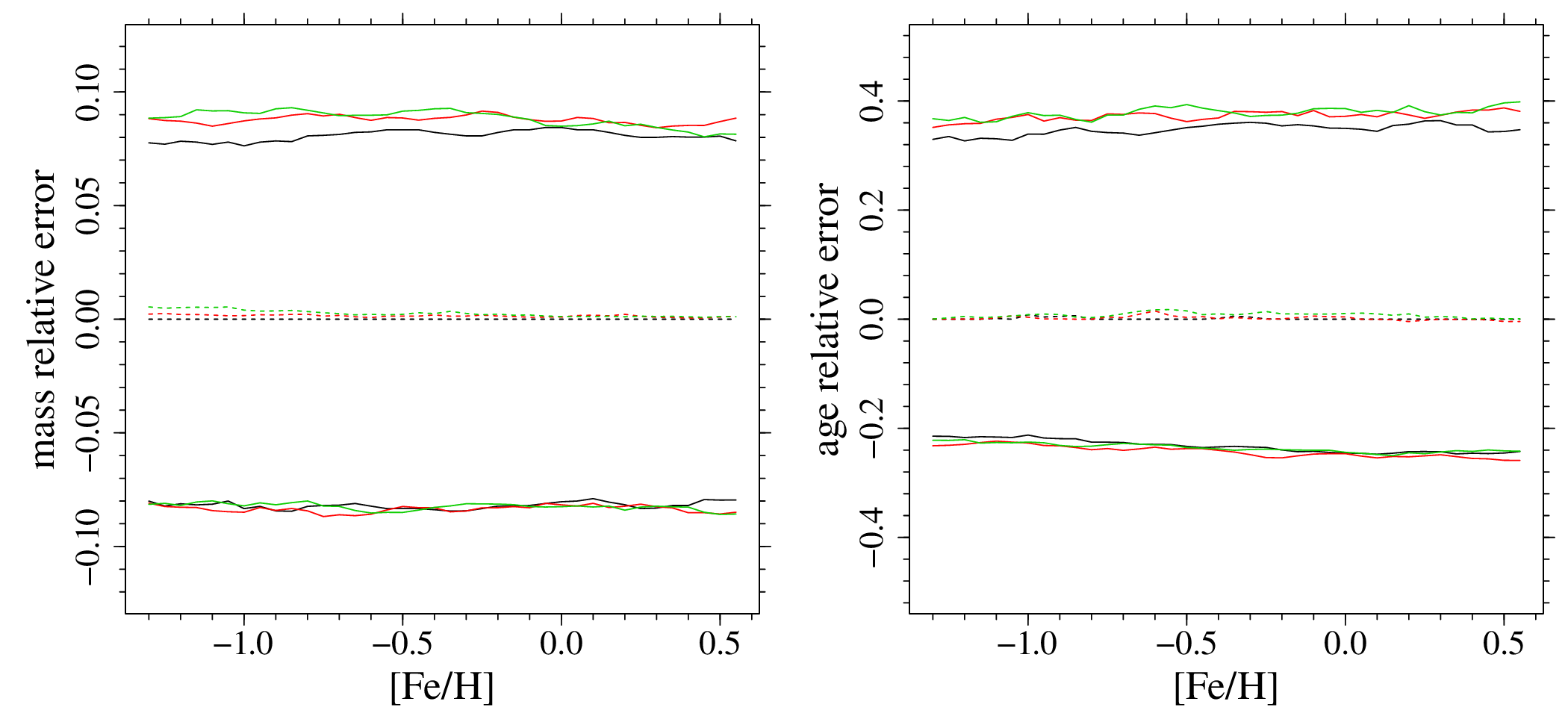}
        \caption{As in Fig.~\ref{fig:envelope-ml}, but as a function of the metallicity [Fe/H].
        }
        \label{fig:envelope-FeH}
\end{figure*}

Figure~\ref{fig:envelope-FeH} shows the dependence of the results on the metallicity of the synthetic stars. The three envelopes evolve at constant width in all the explored range, supporting the conclusion that the impact of the mass loss indetermination does not depend in any way on the metallicity [Fe/H] of the stars, at least in range explored in this work.

\begin{table}[ht]
        \centering
        \caption{Median ($q_{50}$) and  $1 \sigma$ random envelope ($q_{16}$,
                $q_{84}$) of the estimated stellar mass, as a function of the true mass of star.} 
        \label{tab:massvsmass}
        \begin{tabular}{rrrrrrr}
                \hline\hline
                & \multicolumn{6}{c}{Mass ($M_{\sun}$)}\\
                & 0.80 & 1.00 & 1.20 & 1.40 & 1.60 & 1.80 \\ 
                \hline
                \multicolumn{7}{c}{No mass loss scenario}\\
                \hline
                $q_{16}$ & -0.02 & -0.07 & -0.08 & -0.09 & -0.09 & -0.09 \\ 
                $q_{50}$ & 0.00 & 0.00 & 0.00 & 0.00 & 0.00 & -0.01 \\ 
                $q_{84}$ & 0.09 & 0.09 & 0.10 & 0.09 & 0.08 & 0.02 \\ 
                \hline
                \multicolumn{7}{c}{$\eta = 0.4$}\\
                \hline          
                $q_{16}$ & -0.04 & -0.08 & -0.09 & -0.09 & -0.09 & -0.09 \\ 
                $q_{50}$ & 0.01 & 0.00 & 0.00 & 0.00 & 0.00 & -0.01 \\ 
                $q_{84}$ & 0.10 & 0.10 & 0.10 & 0.10 & 0.08 & 0.02 \\ 
                \hline
                \multicolumn{7}{c}{$\eta = 0.8$}\\
                \hline
                $q_{16}$ & -0.04 & -0.07 & -0.09 & -0.09 & -0.09 & -0.09 \\ 
                $q_{50}$ & 0.01 & 0.00 & 0.00 & 0.00 & 0.00 & -0.01 \\ 
                $q_{84}$ & 0.10 & 0.09 & 0.10 & 0.10 & 0.08 & 0.02 \\ 
                \hline
        \end{tabular}
\end{table}

\begin{table}[ht]
        \centering
        \caption{Median ($q_{50}$) and  $1 \sigma$ random envelope ($q_{16}$,
                $q_{84}$) of the relative error on the estimated stellar radius, as a function of the true mass of star.} 
        \label{tab:radiusvsmass}
        \begin{tabular}{rrrrrrr}
                \hline\hline
                & \multicolumn{6}{c}{Mass ($M_{\sun}$)}\\
                & 0.80 & 1.00 & 1.20 & 1.40 & 1.60 & 1.80 \\ 
                \hline
                \multicolumn{7}{c}{No mass loss scenario}\\
                \hline
                $q_{16}$ & -0.01 & -0.03 & -0.04 & -0.04 & -0.04 & -0.04 \\ 
                $q_{50}$ & 0.00 & 0.00 & 0.00 & 0.00 & 0.00 & -0.01 \\ 
                $q_{84}$ & 0.03 & 0.04 & 0.04 & 0.04 & 0.03 & 0.01 \\ 
                \hline
                \multicolumn{7}{c}{$\eta = 0.4$}\\
                \hline
                $q_{16}$ & -0.02 & -0.03 & -0.04 & -0.04 & -0.04 & -0.04 \\ 
                $q_{50}$ & 0.01 & 0.00 & 0.00 & 0.00 & 0.00 & -0.00 \\ 
                $q_{84}$ & 0.04 & 0.04 & 0.04 & 0.04 & 0.03 & 0.01 \\ 
                \hline
                \multicolumn{7}{c}{$\eta = 0.8$}\\
                \hline
                $q_{16}$& -0.02 & -0.03 & -0.04 & -0.04 & -0.04 & -0.04 \\ 
                $q_{50}$& 0.01 & 0.00 & 0.00 & 0.00 & 0.00 & -0.00 \\ 
                $q_{84}$ & 0.04 & 0.04 & 0.04 & 0.04 & 0.03 & 0.01 \\ 
                \hline
        \end{tabular}
\end{table}

\begin{table}[ht]
        \centering
        \caption{Median ($q_{50}$) and  $1 \sigma$ random envelope ($q_{16}$,
                $q_{84}$) of the relative error on the estimated stellar age, as a function of the true mass of star.} 
        \label{tab:agevsmass}
        \begin{tabular}{rrrrrrr}
                \hline\hline
                & \multicolumn{6}{c}{Mass ($M_{\sun}$)}\\
                & 0.80 & 1.00 & 1.20 & 1.40 & 1.60 & 1.80 \\ 
                \hline
                \multicolumn{7}{c}{No mass loss scenario}\\
                \hline
                $q_{16}$ & -0.25 & -0.26 & -0.28 & -0.27 & -0.22 & -0.07 \\ 
                $q_{50}$ & -0.03 & 0.00 & 0.00 & 0.00 & 0.00 & 0.04 \\ 
                $q_{84}$ & 0.12 & 0.29 & 0.38 & 0.37 & 0.37 & 0.36 \\ 
                \hline
                \multicolumn{7}{c}{$\eta = 0.4$}\\
                \hline
                $q_{16}$ & -0.26 & -0.27 & -0.28 & -0.27 & -0.21 & -0.06 \\ 
                $q_{50}$ & -0.03 & 0.00 & 0.00 & 0.00 & 0.00 & 0.03 \\ 
                $q_{84}$ & 0.16 & 0.34 & 0.42 & 0.41 & 0.39 & 0.37 \\ 
                \hline
                \multicolumn{7}{c}{$\eta = 0.8$}\\
                \hline
                $q_{16}$ & -0.25 & -0.26 & -0.27 & -0.27 & -0.21 & -0.06 \\ 
                $q_{50}$ & -0.02 & 0.00 & 0.02 & 0.01 & 0.02 & 0.04 \\ 
                $q_{84}$ & 0.16 & 0.35 & 0.44 & 0.42 & 0.39 & 0.39 \\ 
                \hline
        \end{tabular}
\end{table}

\begin{table}[ht]
        \centering
        \caption{Median ($q_{50}$) and  $1 \sigma$ random envelope ($q_{16}$,
                $q_{84}$) of the relative error on the estimated stellar mass, as a function of the  RGB relative age.} 
        \label{tab:massvsage}
        \begin{tabular}{rrrrrrr}
                \hline\hline
                & \multicolumn{5}{c}{$r$}\\
                & 0.2 & 0.4 & 0.6 & 0.8 & 1.0 \\ 
                \hline
                \multicolumn{6}{c}{No mass loss scenario}\\
                \hline
                $q_{16}$ & -0.05 & -0.08 & -0.09 & -0.08 & -0.07 \\ 
                $q_{50}$ & 0.00 & 0.00 & 0.00 & 0.00 & 0.00 \\ 
                $q_{84}$ & 0.04 & 0.07 & 0.09 & 0.09 & 0.07 \\ 
                \hline
                \multicolumn{6}{c}{$\eta = 0.4$}\\
                \hline
                $q_{16}$ & -0.04 & -0.08 & -0.09 & -0.08 & -0.08 \\ 
                $q_{50}$ & 0.00 & 0.00 & 0.00 & 0.00 & 0.01 \\ 
                $q_{84}$ & 0.05 & 0.07 & 0.09 & 0.09 & 0.12 \\ 
                \hline
                \multicolumn{6}{c}{$\eta = 0.8$}\\
                \hline
                $q_{16}$ & -0.04 & -0.08 & -0.08 & -0.08 & -0.08 \\ 
                $q_{50}$ & 0.00 & 0.00 & 0.00 & 0.00 & 0.02 \\ 
                $q_{84}$ & 0.05 & 0.07 & 0.09 & 0.09 & 0.12 \\ 
                \hline
        \end{tabular}
\end{table}

\begin{table}[ht]
        \centering
        \caption{Median ($q_{50}$) and  $1 \sigma$ random envelope ($q_{16}$,
                $q_{84}$) of the relative error on the estimated stellar radius, as a function of the  RGB relative age.} 
        \label{tab:radiusvsage}
        \begin{tabular}{rrrrrrr}
                \hline\hline
                & \multicolumn{5}{c}{$r$}\\
                & 0.2 & 0.4 & 0.6 & 0.8 & 1.0 \\ 
                \hline
                \multicolumn{6}{c}{No mass loss scenario}\\
                \hline
                $q_{16}$ & -0.02 & -0.03 & -0.04 & -0.04 & -0.03 \\ 
                $q_{50}$ & 0.00 & 0.00 & 0.00 & 0.00 & 0.00 \\ 
                $q_{84}$ & 0.02 & 0.03 & 0.03 & 0.04 & 0.03 \\ 
                \hline
                \multicolumn{6}{c}{$\eta = 0.4$}\\
                \hline  
                $q_{16}$ & -0.02 & -0.03 & -0.04 & -0.04 & -0.04 \\ 
                $q_{50}$ & 0.00 & 0.00 & 0.00 & 0.00 & 0.01 \\ 
                $q_{84}$ & 0.02 & 0.03 & 0.03 & 0.04 & 0.05 \\ 
                \hline
                \multicolumn{6}{c}{$\eta = 0.8$}\\
                \hline  
                $q_{16}$ & -0.02 & -0.03 & -0.04 & -0.04 & -0.04 \\ 
                $q_{50}$ & 0.00 & 0.00 & 0.00 & 0.00 & 0.01 \\ 
                $q_{84}$ & 0.02 & 0.03 & 0.04 & 0.04 & 0.05 \\ 
                \hline
        \end{tabular}
\end{table}

\begin{table}[ht]
        \centering
        \caption{Median ($q_{50}$) and  $1 \sigma$ random envelope ($q_{16}$,
                $q_{84}$) of the relative error on the estimated stellar age, as a function of the  RGB relative age.} 
        \label{tab:agevsage}
        \begin{tabular}{rrrrrrr}
                \hline\hline
                & \multicolumn{5}{c}{$r$}\\
                & 0.2 & 0.4 & 0.6 & 0.8 & 1.0 \\ 
                \hline
                \multicolumn{6}{c}{No mass loss scenario}\\
                \hline
                $q_{16}$ & -0.12 & -0.20 & -0.25 & -0.25 & -0.21 \\ 
                $q_{50}$ & 0.00 & 0.01 & 0.00 & 0.00 & 0.00 \\ 
                $q_{84}$ & 0.16 & 0.35 & 0.37 & 0.36 & 0.27 \\ 
                \hline
                \multicolumn{6}{c}{$\eta = 0.4$}\\
                \hline
                $q_{16}$ & -0.14 & -0.20 & -0.25 & -0.25 & -0.28 \\ 
                $q_{50}$ & -0.01 & 0.00 & 0.00 & 0.00 & 0.00 \\ 
                $q_{84}$ & 0.13 & 0.35 & 0.38 & 0.37 & 0.42 \\ 
                \hline
                \multicolumn{6}{c}{$\eta = 0.8$}\\
                \hline
                $q_{16}$ & -0.13 & -0.21 & -0.25 & -0.24 & -0.25 \\ 
                $q_{50}$ & -0.01 & 0.00 & 0.00 & 0.01 & 0.04 \\ 
                $q_{84}$ & 0.15 & 0.34 & 0.37 & 0.38 & 0.50 \\ 
                \hline
        \end{tabular}
\end{table}

\section{Conclusions}\label{sec:conclusions}

In the framework of the analysis of systematic bias in the estimate of stellar characteristics through grid based techniques adopting asteroseismic constraints, 
we theoretically studied the impact of a wrong assumption about the mass loss efficiency on the estimates of mass, radius, and age for low-mass stars in the RGB phase.      

To this purpose, we used  our grid-based pipeline SCEPtER \citep{scepter1,eta}.  
As observational constraints,
we adopted the stellar effective temperature, its metallicity
[Fe/H], the large frequency spacing $\Delta \nu$, and the frequency of maximum
oscillation power $\nu_{\rm max}$ of the star. The grid of stellar models covers
 the evolutionary phases from ZAMS to the tip of the RGB in
the mass range [0.8; 1.8] $M_{\sun}$ and metallicities [Fe/H] from $-0.55$ to 0.55 dex.

The statistical relative errors on mass and age determinations -- owing to the observational uncertainties alone -- resulted about 8\% and 30\% in the RGB phase, with an increase of almost a factor of three with respect to that at the end of the central hydrogen burning phase. 

We quantified the 
systematic biases due to the uncertainties in the mass loss parameters $\eta$, adopting two possible values of 0.4 and 0.8. 
The most of RGB mass loss happens after the RGB bump phase. Thus for stars up to this evolutionary stage the adoption of a  wrong mass loss efficiency in the recovery has a minor influence on the results, with negligible biases. We also detected a shift of the $1 \sigma$ error envelope of few percent in the last 10\% of the RGB evolution. In the 2.5\% terminal part of the RGB ($\Delta \nu$ under about 1.5 $\mu$Hz) biases are more important, being as larger than random error component for mass determinations, and equal to the random error component for age estimates. These results are insensitive to the stellar metallicity [Fe/H].

The results of the Monte Carlo simulations presented in this paper fully support the conclusion of a very limited impact of a wrong mass loss assumption for the vast majority of stars -- at least in the explored range of $\eta$ values -- on the estimates of stellar parameters by means of grid based technique relying on asteroseismic observables and as long as stars are not too close to the tip of the RGB. This finding is in good agreement with the conclusion by \citet{Casagrande2014, Casagrande2016} who report negligible influence of the mass loss for RGB stars from the analysis of the SAGA sample in the {\it Kepler} field.

\begin{acknowledgements}
We thank our anonymous referee for the useful comments that helped in improving and clarifying the paper.
\end{acknowledgements}

\bibliographystyle{aa}
\bibliography{biblio}

\end{document}